\documentclass[conference]{IEEEtran}
\IEEEoverridecommandlockouts

\usepackage{cite}
\usepackage{amsmath,amssymb,amsfonts}
\usepackage{algorithmic}
\usepackage{graphicx}
\usepackage{textcomp}
\usepackage{ulem}
\usepackage{xcolor}
\usepackage{subcaption}
\usepackage{float}
\usepackage{multirow}
\def\BibTeX{{\rm B\kern-.05em{\sc i\kern-.025em b}\kern-.08em
    T\kern-.1667em\lower.7ex\hbox{E}\kern-.125emX}}

\newcommand{\tup}[1]{\left< #1 \right>}			

\newcommand{\chg}[1]{#1}

\newcommand{\K}{\mathcal{K}}

\title{Malicious Cyber Activity Detection Using \\ Zigzag Persistence
\thanks{This work was supported by the Pacific Northwest National Laboratory operated for the U. S. Department of Energy (DOE) by Battelle under Contract DE-AC06-76RL01830. U.S. Government work not protected by U.S. copyright.}
}

\author{
\IEEEauthorblockN{Audun Myers, Alyson Bittner, Sinan Aksoy, Daniel M. Best, Gregory Henselman-Petrusek,\\
Helen Jenne, Cliff Joslyn, Bill Kay, Garret Seppala, Stephen J. Young, Emilie Purvine}
\IEEEauthorblockA{\textit{Pacific Northwest National Laboratory}\\
Richland and Seattle, WA, United States \\
$\{$audun.myers, alyson.bittner, sinan.aksoy, daniel.best, gregory.roek, helen.jenne,\\ cliff.joslyn, william.kay, garret.seppala, stephen.young, emilie.purvine$\}$@pnnl.gov}
}

\begin{document}

\maketitle


\begin{abstract}

\label{sec:abstract}

In this study we synthesize zigzag persistence from topological data analysis with autoencoder-based approaches to detect malicious cyber activity and derive analytic insights.
Cybersecurity aims to safeguard computers, networks, and servers from various forms of malicious attacks, including network damage, data theft, and activity monitoring. 
Here we focus on the detection of malicious activity using log data. 
To do this we consider the dynamics of the data by exploring the changing topology of a hypergraph representation gaining insights into the underlying activity.
\chg{Hypergraphs provide a natural representation of cyber log data by capturing complex interactions between processes}. To study the changing topology we use zigzag persistence which captures how topological features persist at multiple dimensions over time. \chg{We observe that the resulting barcodes represent malicious activity differently than benign activity.} To automate this detection we implement an autoencoder trained on a vectorization of the resulting zigzag persistence barcodes. Our experimental results demonstrate the effectiveness of the autoencoder in detecting malicious activity \chg{in comparison to standard summary statistics}. Overall, this study highlights the potential of zigzag persistence and its combination with temporal hypergraphs for analyzing cybersecurity log data and detecting malicious behavior.

\end{abstract}


\section{Introduction} \label{sec:introduction}
In this study, we leverage zigzag persistence~\cite{Carlsson2010}, a method from topological data analysis (TDA)~\cite{Edelsbrunner2002,Zomorodian2004}, \chg{coupled with autoencoder anomaly detection} to delve into the temporal activity of cyber data and effectively detect malicious behavior.

Cybersecurity aims to safeguard computers, networks, and \chg{users} from various forms of malicious attacks \chg{that undermine confidentiality, integrity, or availability~\cite{Anderson1972,whitman2021}}. These attacks are typically carried out by gaining unauthorized access \chg{to systems or services}, often \chg{leaving behind} evidence of the attacks in the underlying log data which captures information such as timestamps, Internet Protocol (IP) addresses, ports, \chg{executable paths, and command line entries}. However, detecting that malicious activity in the log data is challenging due to \chg{the data's} size and complexity. 

One common approach to finding malicious activity in cyber logs involves constructing \chg{and analyzing} graph representations of the data, such as process trees \cite{Mamun2021} or flow networks \cite{aksoy2021directional}, that model dyadic relations between entities. However, standard graphs cannot capture multi-way interactions that are common in cyber data. Instead, using higher dimensional graphs, known as hypergraphs~\cite{berge1984hypergraphs}, for modeling cyber log data more effectively captures the complex interactions present between users, processes, ports, and other resources. Hypergraphs have proven valuable in diverse branches of data science, including machine learning, biology, and social networks \cite{han1998hypergraph,Klamt2009cellular,zlatic2009hypergraph}.

While hypergraphs capture the complex multi-way relationships, traditional {\it static} hypergraphs may fail to additionally represent the dynamic nature of cyber systems. By incorporating temporal information on vertices, hyperedges, or incidences, \textit{temporal hypergraphs}~\cite{neuhauser2021consensus, fischer2020visual, antelmi2020design} offer a solution. \chg{Temporal hypergraphs allow vertices and hyperedges} to appear and disappear over time and connect different sets of \chg{vertices} at different points in time. As such, they provide a suitable framework for studying dynamical systems of complex relations. Cyber log data falls squarely into this category as temporal information is present on all log records, and each record or collection of records captures complex relationships among groups of network entities, e.g., hosts, IPs, ports, users, and executable files.

One approach to \chg{representing} temporal hypergraphs that we implement in this work is as a sequence, one hypergraph per sliding time window, representing the state of the system \chg{during that time}. 
This sequential representation of the temporal hypergraph allows one to treat the sequence as a dynamical process $G_t \mapsto G_{t+1}$ gaining a dynamical systems perspective.
Each hypergraph in the sequence is a set of \chg{vertices} and a multiset of hyperedges. \chg{Each vertex and hyperedge represents a distinct named entity (e.g., an IP, port, user, program executable). All vertices are the same type (e.g., all IPs), as are all edges, but edges and vertices represent different types. The hyperedge corresponding to a specific entity can include different sets of vertices at different times. This will be made more concrete in Section \ref{sec:background}.}

Our primary objective is to analyze temporal hypergraph representations of cyber log data to effectively detect malicious activity. 
Our claim is that malicious cyber activity will often exhibit unique attack patterns in the log data, resulting in topological changes in the representations over time. 
Specifically, we investigate a hypergraph representation constructed with executables as \chg{vertices} and destination ports as hyperedges.
\chg{The dynamics of the topology of this hypergraph should be different during malicious times than it is during benign since} malicious activity often has more complexity and quantity in the executables interactions changing at faster time scales compared to benign activity.
This intuition will be illustrated in Section~\ref{sec:intuition}.
\chg{We considered many combinations of hyperedges and vertices for constructing hypergraphs and found the clearest malicious activity detection using this construction.}

The evolution of hypergraph structure and topology over time naturally fits into a use case of zigzag persistence, a tool from TDA.
With temporal hypergraphs providing a valuable framework for capturing complex dynamical systems we need to build an understanding of the complex patterns and structural changes in these temporal hypergraphs, and this is where zigzag persistence comes into play. Zigzag persistence captures how, when, and for how long topological features at multiple dimensions persist. For example, is a \chg{distinct} component \chg{seen over} a long time, and \chg{if so, is it} always present, or does it intermittently appear? 

Zigzag persistence has been previously used for studying temporal graph models~\cite{Myers2022a} of transportation networks and for intermittency detection. This method has also been recently extended to study temporal hypergraphs for both cyber and social network data~\cite{myers2023topological}.
By leveraging the power of zigzag persistence, one is able to delve deep into the intricate temporal dynamics of (hyper)graphs, unveiling hidden trends, detecting critical events, and revealing the underlying structural transformations that shape the system's behavior. 

To determine the viability of this approach we implement an autoencoder as a form of anomaly detection on a vectorization of the resulting zigzag persistence barcodes. We train the model to detect suspicious activity and investigate vectors that have high reconstruction loss.
\chg{We chose to use an autoencoder based on the assumption that a large proportion of traffic on the network is typical benign activity, whereas malicious activity is fairly uncommon.}


We begin in Section~\ref{sec:background} by introducing notation and definitions for temporal hypergraphs, zigzag persistence, and how we use zigzag persistence to study temporal hypergraphs. We also introduce the concept of an autoencoder. 
In Section~\ref{sec:methods} we describe the cyber data, our experimental design, and some intuition behind using dynamic topology to identify anomalous behavior. We then demonstrate the ability of the pipeline to detect malicious activity in Section \ref{sec:results}. We provide future goals and conclusions on this work in Section~\ref{sec:conclusion}.

\section{Computational Tools} \label{sec:background}

The process of computing zigzag persistence for a temporal hypergraph begins with a sequence of representative hypergraphs. We then transform each hypergraph into an abstract simplicial complex and examine the appearance and disappearance of topological features across multiple dimensions in this sequence using zigzag persistence.
In the final step of our pipeline we vectorize the zigzag persistence barcode and use an autoencoder to identify anomalous barcodes.
In order to describe our experimental design in the context of cyber log data in Section~\ref{ssec:experimental_design}, we begin by first introducing the necessary definitions and background in a general setting.

\subsection{Hypergraphs and Abstract Simplicial Complexes}
A \emph{hypergraph}, $G=(V, E)$, analogous to a graph, is represented by a set of \emph{vertices}, $V$ and a \chg{family} of \emph{(hyper)edges} $E$. The main difference between a hypergraph and a classical graph is that an edge $e \in E$ can be an arbitrary subset of vertices $e \subseteq V$ as opposed to a pair. If $|e|=k$ then we say that $e$ is a \emph{$k$-edge}.
A \emph{temporal hypergraph} is a sequence of $n$ hypergraphs, denoted as $\mathcal{G} = G_0, G_1, G_2, \ldots, G_{n-1}$, \chg{where $G_i = \tup{V,E_i}$}. The sequence can be viewed as a discrete dynamical process, where $G_t$ transitions to $G_{t+1}$, enabling us to gain insights into the dynamics of the underlying system.

An \emph{abstract simplicial complex} (ASC), denoted as $K$, is a \chg{non-empty} collection of \chg{non-empty} sets that is closed under taking subsets. Formally, $K=\{\sigma\}$ is an ASC if whenever $\tau \subset \sigma \in K$ then $\tau \in K$. Each set, $\sigma$, is called a \emph{simplex}, and if $|\sigma|=k$ then $\sigma$ has dimension $k-1$ and is called a $(k-1)$-simplex. 
Geometrically, 0-simplices represent points or vertices, 1-simplices represent lines or edges, 2-simplices represent filled-in triangles, \chg{3-simplices filled in tetrahedrons, and so on for arbitrary hyper-tetrahedrons (see Figure \ref{fig:feature_dying})}. For $\tau, \sigma \in K$ we say that $\tau\neq\emptyset$ is a face of $\sigma$ if $\tau \subseteq \sigma$. The definition of an ASC implies that every simplex is closed under the face relation, meaning it includes all of its faces (except for the empty set) as defined by the power set of the simplex.

Note that an ASC can be thought of as a hypergraph with an extra requirement on the edges, but the reverse is not true: a general hypergraph need not be an ASC. Although various methods exist for constructing an ASC from a hypergraph~\cite{Gasparovic2021} in this paper we consider the \emph{associated ASC of a hypergraph}~\cite{Ren2020}. The associated ASC consists of a simplex for each hyperedge. In other words, the associated ASC of a hypergraph $G$ contains all subsets of all hyperedges:
\[ K(G) = \{ \sigma \subseteq V : \exists e \in E, \sigma \subseteq e \}.\]
As many real-world hypergraphs have some large hyperedges, \chg{constructing $K(G)$}  can be costly, and unnecessary if computing only low dimensional homology.
In practice, to reduce computational complexity, we keep only those simplicies up to a small maximum dimension, $p=2$ or $3$. 

\subsection{Simplicial Homology} \label{ssec:simplicial_homology}
\emph{Simplicial homology} is an algebraic approach to analyze the structure of an ASC by quantifying the number of $p$-dimensional \emph{features}. 0-dimensional features are connected components, 1-dimensional features are graph cycles, 2-dimensional features are hollow tetrahedra, and so on.
The $p$-dimensional simplicial homology of an ASC, $K$, denoted $H_p(K)$, is a vector space whose basis represents the $p$-dimensional features of $K$. The rank of $H_p(K)$ then counts the number of $p$-dimensional features. This rank is denoted $\beta_p$ and called the $p^{th}$ \emph{Betti number} of $K$.
The algebraic details of simplicial homology computations and Betti numbers can be found in \cite{hatcher}.

While Betti numbers provide valuable insights into the changing topology of hypergraph snapshots, they do not capture the relationships between the topology of consecutive snapshots. In other words, Betti numbers alone do not reveal if a feature persists throughout the entire sequence. To address this limitation and track the changes in homology and their interconnections across a sequence of ASCs, we employ the technique of zigzag persistent homology.

\subsection{Persistent and Zigzag Homology} \label{ssec:ZZ_THG}
This section provides an introduction to persistent homology (PH)~\cite{Zomorodian2004} and how it generalizes to zigzag persistent homology. For a detailed introduction to PH we suggest \cite{Otter2017,Munch2017}, for zigzag see~\cite{Carlsson2010}.

PH is used to obtain a sense of the shape and size of a data set at multiple scale resolutions. To gain some intuition on what this means we describe a common setting in which PH is applied, that of a point cloud $X \subseteq \mathbb{R}^n$. At a given scale (i.e., distance value) we connect points in $X$ within the given distance to form an ASC. As that scale increases so does the ASC and topological features are born (appear) and die (are filled in). PH tracks the birth and death of these features as the distance scale varies to form a topological fingerprint. Short-lived features may indicate noise while long-lived ones often indicate  meaningful features. The birth and death thresholds provide an idea of the general size or geometry of each feature, which can in turn provide intuition and interpretation back into the data itself. For example, the presence of a 1-dimensional loop might mean that the data is cyclical or repetitive whereas the presence of multiple 0-dimensional components could indicate strong clustering of the data. 

\begin{figure}[t]
    \centering
    \includegraphics[width=0.7\linewidth]{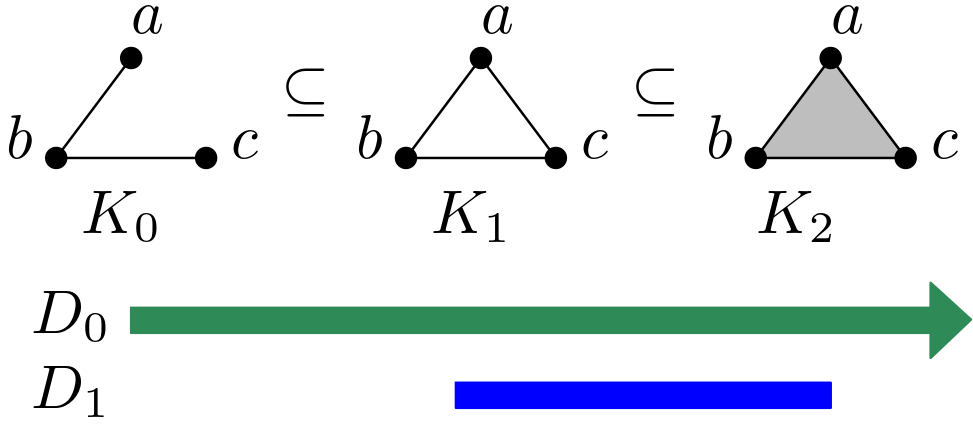}
    \caption{\chg{Example of a nested zigzag sequence of ASCs. (Left) Two graph edges (2-simplices) forming a graph chain. (Center) Three graph edges (2-simplices) forming a graph cycle. (Right) A 2-simplex (filled triangle). The $D_0$ and $D_1$ PH barcodes are shown below.}}
    \label{fig:feature_dying}
\end{figure}

A point cloud is not the only setting for PH. In general, only a sequence of nested ASCs\footnote{In fact, persistent homology can be applied in even more general settings but for the purposes of this paper we won't consider arbitrary topological spaces or chain complexes.}, often referred to as a \emph{filtration}, is necessary:
\begin{equation}
\K = K_{0} \subseteq K_{1} \subseteq K_{2} \subseteq \ldots \subseteq K_{n}.
\label{eq:nested_complexes}
\end{equation}
For a given dimension $p$ we can calculate $H_p(K_i)$ for each $K_i$.
In order to capture how the homology changes from $K_i$ to $K_{i+1}$ we rely on the fact that $K_i$ is a sub-complex of $K_{i+1}$ and so the components of the topological features found in $K_i$ (e.g., the vertices, edges, and higher dimensional simplices) must also be found in $K_{i+1}$. If these components also form a topological feature in $K_{i+1}$ then the feature \emph{persists}. If they do not form a feature in $K_{i+1}$ then the feature \emph{dies}. In Figure~\ref{fig:feature_dying} we see a \chg{filtration with a} 1-dimensional feature in $K_1$ consisting of the edges $(a,b), (a,c), (b,c)$. These edges are present in $K_{2}$ but they no longer form a 1-dimensional feature because of the presence of the triangle $(a,b,c)$.
The appearance and disappearance of $p$-dimensional features in the filtration is tracked in a summary known as a persistence \emph{barcode}, a collection of intervals, one for each topological feature identified.
Each feature has an associated interval $[b, d]$ that indicates the index of the appearance of the feature, its \emph{birth} threshold $b$, and its disappearance, its \emph{death} threshold $d$.
\chg{If a feature is present in the final ASC in the sequence we say its death is $\infty$ because it does not die within the filtration.}
We denote the barcode for dimension $p$ of a sequence $\K$ as $D_p(\K)=\{[b_i, d_i]\}$, or simply $D_p$ is the sequence is clear from context.
In the example in Figure~\ref{fig:feature_dying} the 1-dimensional feature is born at $b=1$ and dies at $d=2$.
The algebraic mechanics of tracking features across spaces via their inclusions is best left to the references cited above. For the purposes of this paper only the intuition is necessary. 

Given a temporal hypergraph sequence we can construct $K_i := K(G_i)$. If we are lucky enough to have a sequence in which $K_i \subseteq K_{i+1}$ for all $i$ then we can apply PH directly. However, this is rarely the case. There are plenty of examples in which hypergraph vertices and edges are both added {\it and} removed over time. This is where zigzag homology, which extends the concept of PH to handle ASC sequences with addition and removal of simplices, can be applied. Given an arbitrary sequence of ASCs, $K_0, K_1, \ldots, K_n$, we can form an augmented sequence with interwoven unions\footnote{\chg{Zigzag persistence is also defined for intersections, with the subset containments flipped. Here we explore only the union case.}}:
\[K_0 ~\subseteq~ K_0 ~\cup~ K_1 ~\supseteq~ K_1 ~\subseteq~ K_1 \cup K_2  \cdots K_{n-1} ~\cup~ K_n \supseteq K_n.\]

\begin{figure}[t]
    \centering
    \includegraphics[width=0.7\linewidth]{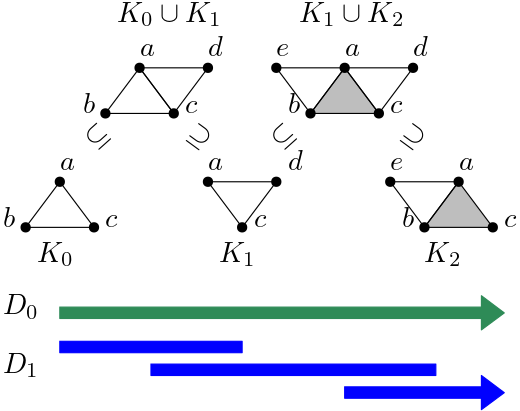}
    \caption{\chg{Example of a zigzag sequence with interwoven unions. The $D_0$ and $D_1$ zigzag persistence barcodes are shown below.}}
    \label{fig:zigzag_ex}
\end{figure}

The idea of zigzag homology is similar to PH. Even though the inclusions are not in the same direction throughout the augmented sequence their presence still allows us to track whether a feature in one ASC is the same as a feature in the next.
In Figure~\ref{fig:zigzag_ex} we show an example sequence of three ASCs with interwoven unions. There is a 1-dimensional feature in all three ASCs but through the use of zigzag we can see that they are all different loops. The barcode consists of three intervals: [0,1] for loop $(a, b), (a, c), (b, c)$, [0.5, 2] for loop $(a, c), (a, d), (c, d)$, and [1.5, $\infty$] for loop $(a, b), (a, e), (b,e)$. If a loop is born (resp. dies) at a union step between $i$ and $i+1$ we say that it is born (resp. dies) at the midpoint, $i+\frac{1}{2}$.

For a more detailed introduction to zigzag persistence in the context of studying temporal hypergraphs, we refer the reader to~\cite{myers2023topological}, which includes an example illustrating the procedure.

\subsection{Vectorization of Persistence Barcodes} \label{ssec:ACCs}
To implement an autoencoder for studying zigzag persistence barcodes we need to create a faithful vector representation of the barcode.
While there are many methods for vectorizing a barcode for machine learning applications, such as persistence images~\cite{Adams2017} and persistence landscapes~\cite{Bubenik2015}, these are often high dimensional making the autoencoder training more burdensome. In this work we use Adcock-Carlsson Coordinates (ACCs)~\cite{adcock2013ring} as they are computationally and storage efficient and have been shown to provide comparable performance to the more advanced vectorization methods for classification tasks~\cite{barnes2021comparative}. The ACCs are calculated as 

\begin{equation} \label{eq:ACC}
    \begin{split}
        ACC(D_p) = \Big[ & \sum_i b_i(d_i - b_i),  \sum_i (d_{\rm max} - d_i)(d_i-b_i),  \\
                         & \sum_i b_i^2(d_i-b_i)^4,  \sum_i (d_{\rm max} - d_i)^2(d_i-b_i)^4 \Big].
    \end{split}
\end{equation}
We then stacked the ACCs for each dimension $p \in [0, 1]$ into a single eight-dimensional vector.

\subsection{Autoencoder}

\begin{figure}[h!]
    \centering
    \includegraphics[width=0.3\textwidth]{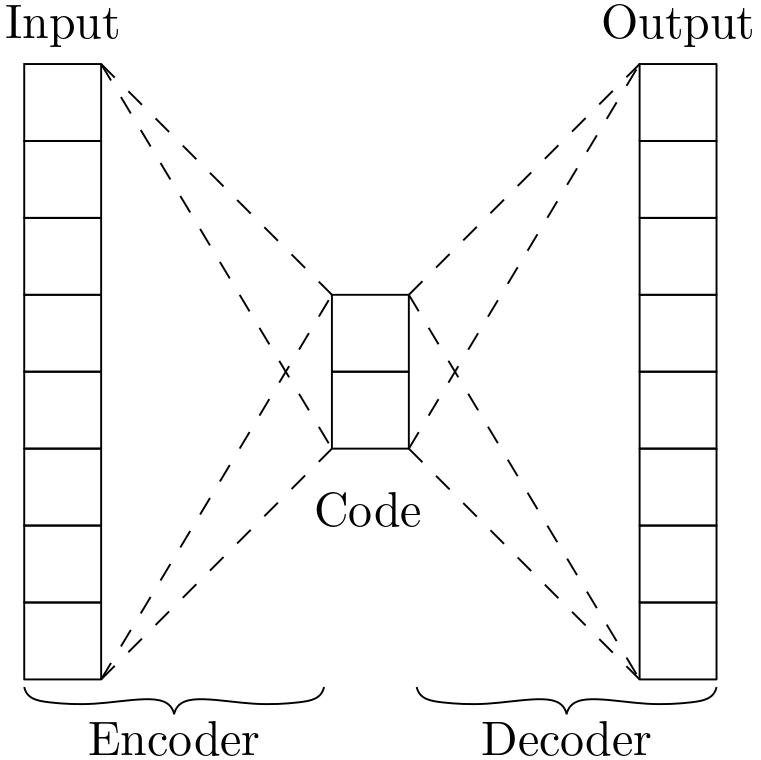}
    \caption{\chg{Our autoencoder schema.}}
    \label{fig:AutoencoderSchema}
\end{figure}

One of the ways to leverage the power of neural networks to perform anomaly detection on a dataset is through the use of autoencoders. An autoencoder is a particular kind of feed-forward neural network that takes in data, compresses it via encoding layers, and then attempts to reconstruct the original representation from the compressed form through decoding layers \chg{as shown in Fig.~\ref{fig:AutoencoderSchema}}. The metric used to quantify the difference between the reconstructed version and the original data is called the reconstruction loss.

If an autoencoder is trained on ``typical" data, then the reconstruction loss for unseen typical data should be low whereas the reconstruction loss for ``atypical" data will be much greater. This is the motivation for utilizing autoencoders to detect anomalies in data. More precisely, if the reconstruction loss of unseen data is above a chosen threshold then the unseen data is considered anomalous.


\section{Methodology}\label{sec:methods}

\subsection{Data and Data Preparation} \label{ssec:data}
The Operationally Transparent Cyber (OpTC) dataset~\cite{OpTC} used in our experiments was created by the Defense Advanced Research Projects Agency (DARPA) as part of a mission to test scaling of cyber attack detection. The data consists of log records of both benign and malicious activity, with an associated ground truth document describing the attack events.
The attack events include downloading \chg{and executing} malicious PowerShell Empire \chg{payloads}, privilege escalation, credential theft, network scanning, and lateral movement.
The data contains both flow and host logs. The elements of each record vary depending on the type of log but the format is standardized allowing for easy analysis across log-types.
In this paper we consider only the flow subset of records \chg{and only 4 of the 58 data fields available. In future work we plan to complete a more comprehensive analysis. The subset of keys in the flow records we use are time, destination port, source IP, and image path (i.e., executable). A sample of records, restricted to those fields, is shown in Table~\ref{tab:OpTC_log_data}.}
\begin{table}[h]
\centering
\chg{\begin{tabular}{llll}
\textbf{Time}                & \multicolumn{1}{c}{\textbf{Dest. Port}} & \multicolumn{1}{c}{\textbf{Source IP}} & \multicolumn{1}{c}{\textbf{Image Path}} \\ \hline
9/23/19 11:25                & 80                                      & 142.20.56.202                          & powershell.exe                          \\
9/23/19 11:25                & 5355                                    & 10.20.1.209                            & svchost.exe                             \\
9/23/19 11:25                & 5355                                    & 10.20.1.209                            & svchost.exe                             \\
9/23/19 11:25                & 5355                                    & 142.20.56.149                          & svchost.exe                             \\
9/23/19 11:25                & 5355                                    & 142.20.56.139                          & svchost.exe                             \\
9/23/19 11:25                & 8000                                    & 142.20.56.202                          & firefox.exe                             \\
9/23/19 11:25                & 5355                                    & 10.20.2.67                             & svchost.exe                             \\
\multicolumn{1}{c}{$\vdots$} & \multicolumn{1}{c}{$\vdots$}            & \multicolumn{1}{c}{$\vdots$}           & \multicolumn{1}{c}{$\vdots$}            \\
9/23/19 11:45                & 138                                     & 142.20.58.104                          & System                                  \\
9/23/19 11:45                & 5355                                    & 10.20.2.164                            & svchost.exe                             \\
9/23/19 11:45                & 5355                                    & 10.20.2.164                            & svchost.exe                             \\ \hline
\end{tabular}}
\caption{\chg{Example OpTC log data of four keys used to construct destination-image path hypergraphs on host 201.}}
\label{tab:OpTC_log_data}
\end{table}

We focus our analysis of the data on the first day of malicious activity, September 23, on a sampling of both benign and malicious hosts, see Table~\ref{tab:hosts}. \chg{Here we classify a host as malicious if there is any malicious activity that occurred on that host, according to the ground truth document.
We chose hosts 201, 402, and 660 as our malicious hosts. The data from these hosts forms our test set.}
For the benign set, we identified hosts that did not appear in the ground truth document and then chose a subset of those hosts with varying levels of activity relative to the malicious hosts. In particular, hosts 0005, 0006, 0010, 0012 had significantly less (approximately half as much) activity, hosts 0162, 0304, 0461, 0906 had comparable amounts of activity, and hosts 0071, 0213, 0222, 0274 had more activity compared to the malicious hosts. 
\chg{Data from the benign hosts forms our training set.}

\begin{table}[h]
    \centering
    \chg{\begin{tabular}{c|l}
        \multirow{2}{*}{Benign (Training) Hosts:} & 0005, 0006, 0010, 0012, 0071, 0162    \\
                     & 0213, 0222, 0274, 0304, 0461, 0906    \\\hline
        \multirow{2}{*}{Malicious (Testing) Hosts:} & \multirow{2}{*}{0201, 0402, 0660}\\
                        &
    \end{tabular}}
    \caption{Subset of hosts used for training and testing}
    \label{tab:hosts}
\end{table}

\begin{figure*}[h]
    \centering
    \includegraphics[width=0.95\linewidth]{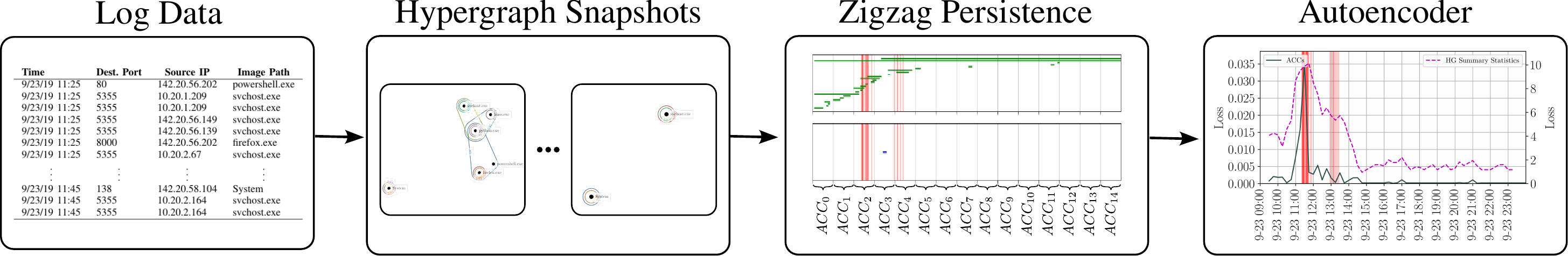}
    \caption{\chg{Experimental design pipeline for study OpTC log data with autoencoder trained on the ACC vectors (e.g., $ACC_0, \ldots, ACC_{14}$) of the subwindowed zigzag persistence barcodes. An additional autoencoder trained on summary statistics of the hypergraph snapshots was also used for comparison without that pipeline shown.} 
    }
    \label{fig:experiment_design_pipeline}
\end{figure*}
We performed selective filtering of the data as an initial preprocessing step. In particular, we filtered out actions where the image path or source IP address were missing and where the source IP address corresponded to local host activity. Since the network traffic data in the dataset is \chg{unidirectional}, we also filtered out actions where the destination port was ephemeral \chg{thereby focusing on flow records where the source IP is the likely originator of the communication. Ephemeral ports, also called dynamic ports, are port numbers above 49152 that are not formally assigned to a service designation and are often used by the originator of the communication.}


\subsection{Experimental Design} \label{ssec:experimental_design}

We designed an  experiment with the aim to identify source IPs that are responsible for malicious activity \chg{captured on a host} and the particular time \chg{window} in which the malicious activity occurs, by using the topology of the interactions \chg{of the IPs with image paths}.
We create hypergraphs for a given source IP and sequence of timeframes, and then vectorize the hypergraph sequences in two ways: 1) using zigzag persistence and 2) a more naive hypergraph property embedding.
In order to understand the viability of zigzag persistence diagrams to encode differences in the topological dynamics of benign and malicious activity we trained two autoencoders, one on the vectors derived from zigzag persistence and a second on the hypergraph property vectors. 
We then perform autoencoder-based anomaly detection separately on the two vectorizations and examine how the anomalies align with the ground truth document. 
If our zigzag autoencoder successfully identifies malicious activity on the network, this provides evidence that the topological information encoded by the zigzag persistence barcodes can aid in cybersecurity efforts.
We use the autoencoder trained on hypergraph property vectors as a comparison.


The details and pipeline of these experiments are illustrated in Fig.~\ref{fig:experiment_design_pipeline}. 
Our experimental design begins with the log data, see the box labeled \textit{Log Data} in Fig.~\ref{fig:experiment_design_pipeline}. We show a small set of the OpTC log data including the specific columns needed: timestamp, source IP, destination port, and image path (executable).
Using the timestamps, we break this log data into 10-minute windows that overlap by 5 minutes.
\chg{We then filter down the 10 minutes of data from each window by a source IP to construct a hypergraph for each window where vertices are the executable files and hyperedges are the destination ports.}
Specifically, for the hypergraph pertaining to source IP $X$ the vertex for executable $t$ is contained in the destination port edge $r$ if there is a record with the (source IP, destination port, executable) tuple $(X, t, r)$.

For each source IP we apply zigzag persistence to the temporal sequence of hypergraph snapshots, as shown in  Fig.~\ref{fig:experiment_design_pipeline} in the box labeled  \textit{Zigzag Persistence}, resulting in a barcode for each dimension (0 and 1). This full time barcode is further broken into sub-barcodes over 1 hour sub-windows. Each of these sub-barcodes are vectorized using the ACCs described in Section~\ref{ssec:ACCs}. 
We trained the zigzag autoencoder on these ACC vectors from IPs in the benign host list from Table~\ref{tab:hosts} and tested on those from the evaluation hosts. \chg{We initialized the autoencoder with random weights using random seed 0.}
For each source IP we calculated the time series of mean squared error reconstruction loss as an indicator of  abnormal or malicious activity. This is shown in Fig.~\ref{fig:experiment_design_pipeline}, in the box labeled \textit{Autoencoder}.


The zigzag autoencoder contains one fully-connected neural layer as the encoder and decoder. The input zigzag vectors are 8-dimensional, the autoencoder compresses the data into 2-dimensional vectors, and decompresses them back into 8-dimensions, \chg{as illustrated in Figure \ref{fig:AutoencoderSchema}. We chose this shallow single layer encoder/decoder schema due to the low dimensionality of the ACC vectors.} The encoder and decoder of the model learn by minimizing the mean squared error between the original vector and the reconstructed \chg{vector}.

We trained a second autoencoder on some \chg{standard} summary statistics of the hypergraphs as a feature vector on the collection of hypergraphs that occurred during the 1 hour sub-windows. For each of the hypergraphs during each sub-window we calculated the number of edges, number of vertices, number of components, and diameter of the largest component and then concatenated them together. This results in a 48-dimensional feature vector for each 1 hour window \chg{that also should capture the dynamics}. The autoencoder again had a latent space of 2-dimensions to make a fair comparison to the first autoencoder.

By analyzing the reconstruction loss of the two autoencoders, we can compare the ability of the zigzag persistence barcodes and standard summary statistics to detect malicious activity.

\subsection{Intuition}\label{sec:intuition}
\begin{figure}[h]
    \centering
    \begin{subfigure}[t]{.24\textwidth}
        \centering
        \includegraphics[width=.95\textwidth]{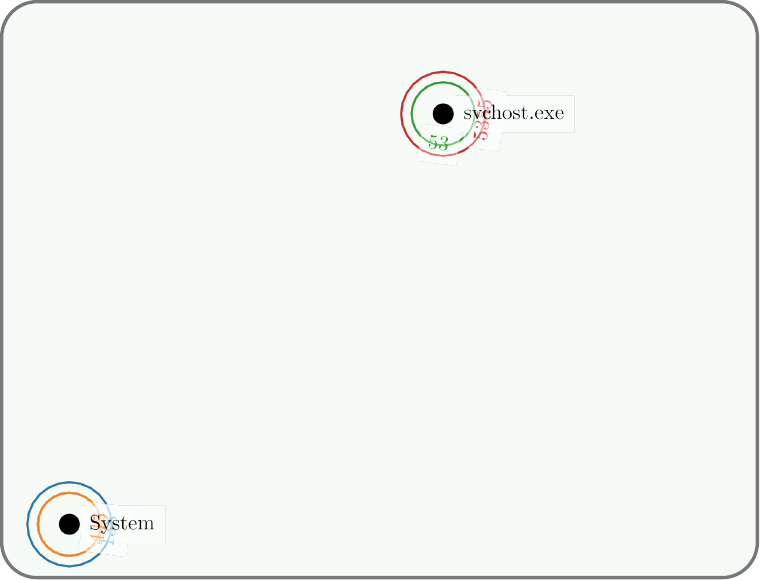}
        \caption{Benign Activity}
        \label{fig:benign_example}
    \end{subfigure}
    \begin{subfigure}[t]{.24\textwidth}
        \centering
        \includegraphics[width=.95\textwidth]{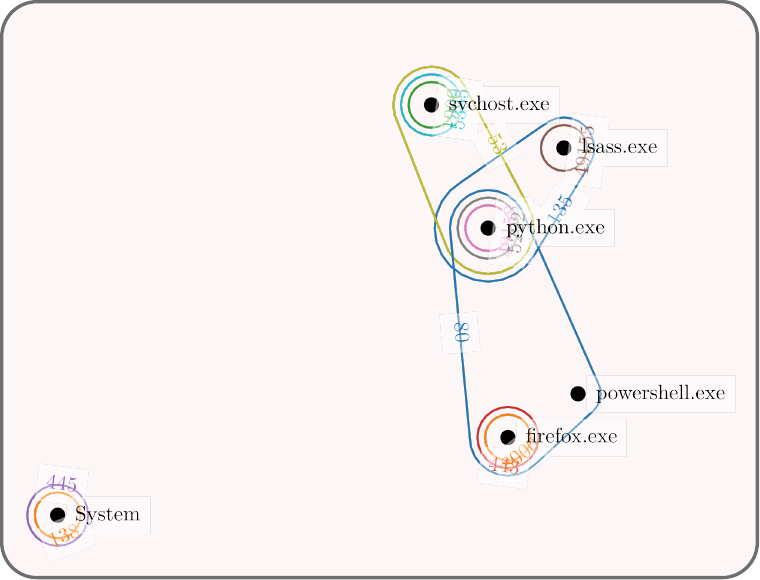}%
        \caption{Malicious Activity.}
        \label{fig:malicious_example}
    \end{subfigure}
    \caption{Hypergraphs formed during malicious and benign activity for source IP 142.20.56.202 on host 201 using destination ports as hyperedges and image path executables as vertices.}
    \label{fig:benign_vs_malicious_activity_snapshots}
\end{figure}

Before we transition to the results of our experiment we provide some intuition, through an example, for why the {\it dynamics} of the hypergraph topology, and not just the static topology of each snapshot, are important for detecting malicious activity.
Fig.~\ref{fig:benign_vs_malicious_activity_snapshots} shows hypergraphs from one benign and one malicious time period for source IP 142.20.56.202 on Host 201. It is apparent that the structural configuration of the hypergraph during benign activity differs from that during malicious activity, but from a topological perspective, the two snapshots are equivalent. Both hypergraphs exhibit two components and no higher-dimensional homology, indicating a similarity in their topological properties.

\begin{figure}[h]
    \centering
    \begin{subfigure}[t]{.46\textwidth}
        \centering
        \includegraphics[width=.99\textwidth]{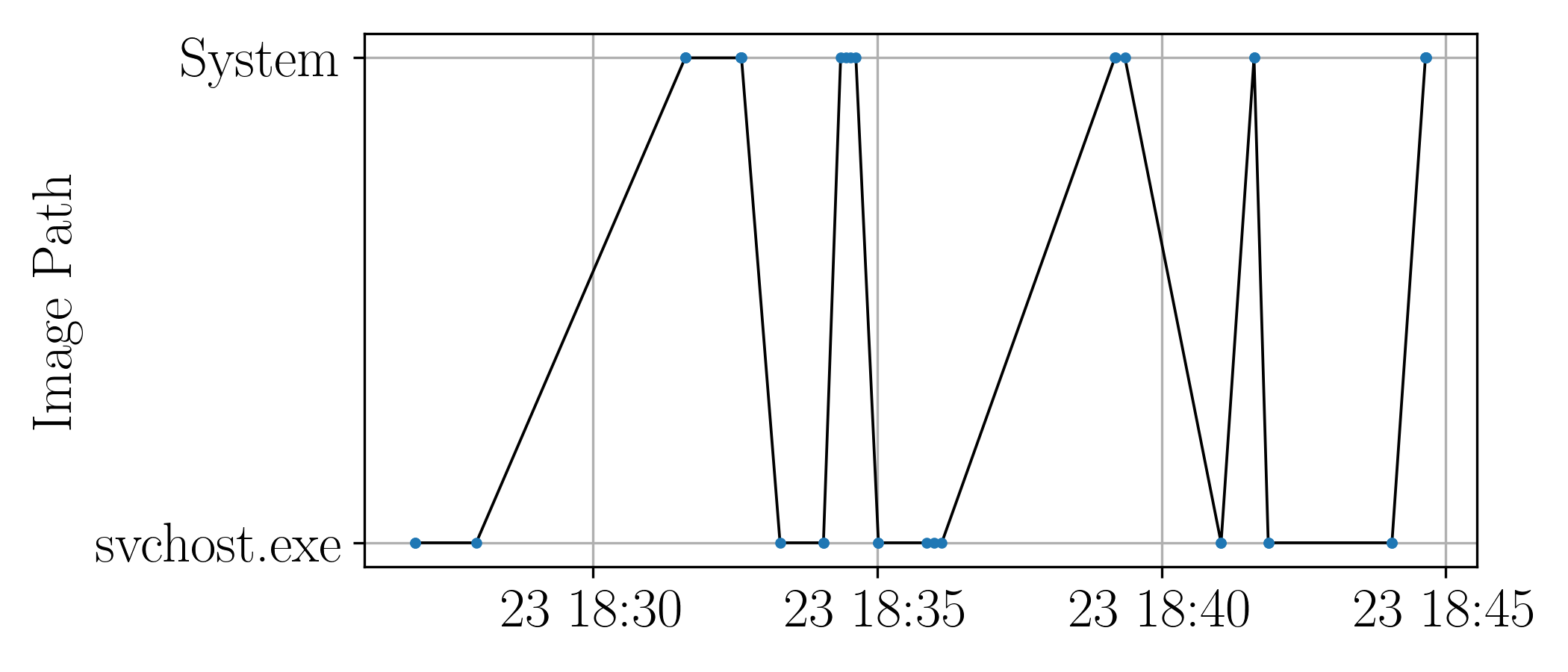}
        \caption{Benign Activity}
        \label{fig:benign_image_path_signal}
    \end{subfigure}
    \vspace{25pt}
    \begin{subfigure}[t]{.46\textwidth}
        \centering
        \includegraphics[width=.99\textwidth]{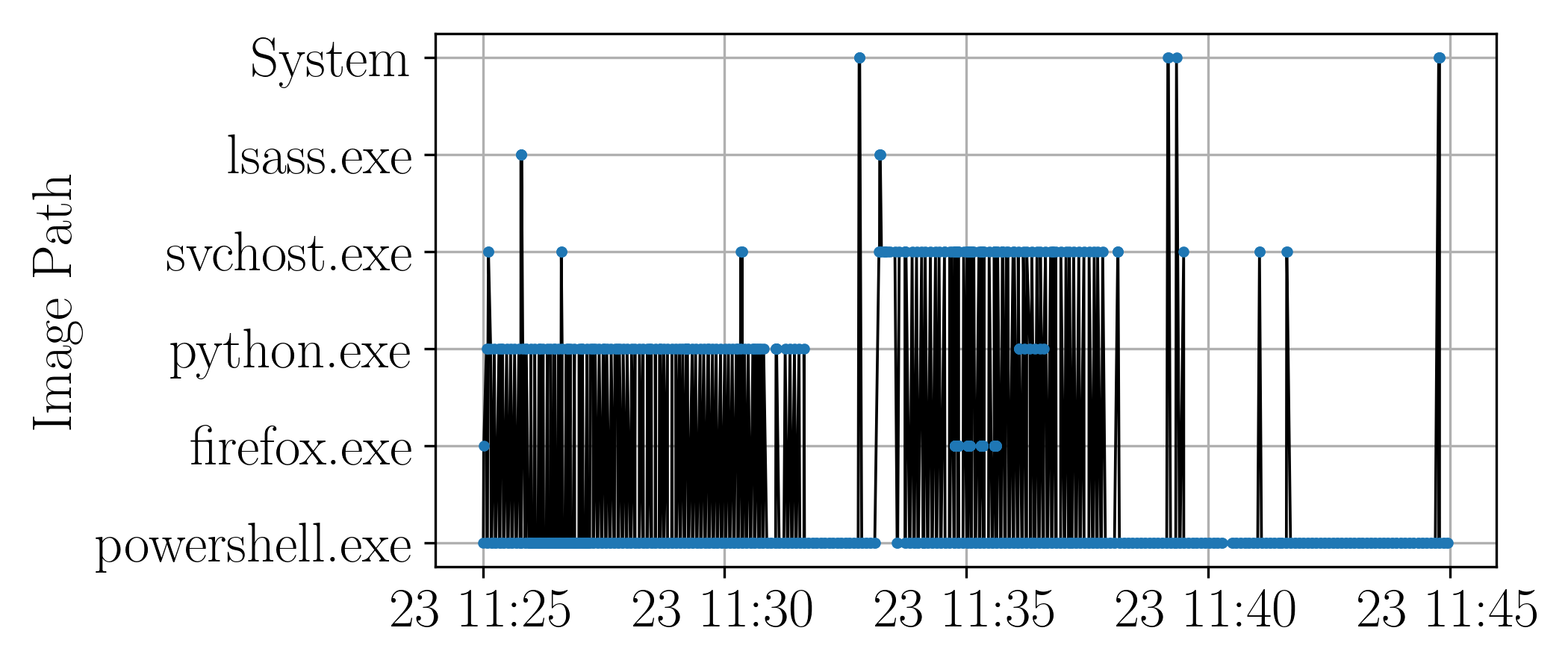}%
        \caption{Malicious Activity.}
        \label{fig:malicious_image_path_signal}
    \end{subfigure}
    \caption{\chg{Sequence of image path executables for source IP 142.20.56.202 during 20 minute benign and malicious activity windows on host 201 (same windows as used to generate the hypergraphs in Fig.~\ref{fig:benign_vs_malicious_activity_snapshots}). }}
    \label{fig:benign_vs_malicious_signals}
\end{figure}

However, the two isolated snapshots do not tell the entire story of the topology. While the snapshots are topologically equivalent they do not account for the underlying dynamics of the topology (e.g., do these two components persist for long periods of time or do they quickly evolve?). 
By looking beyond the isolated snapshots we can gain a quick insight that, in fact, the dynamics of the malicious activity change at a much higher rate than the dynamics of the benign activity. Figure~\ref{fig:benign_vs_malicious_signals} shows the sequence of image path (executables) for source IP 142.20.56.202 on host 201 during the same 20 minute benign and malicious activity windows associated to the hypergraphs in Fig.~\ref{fig:benign_vs_malicious_activity_snapshots}. As shown, during the benign activity the only executables used were System and svchost.exe and tend to be executed every few minutes, while in the malicious activity many executables are used and are executed much more frequently.  \chg{During the benign activity the processes are associated with operating system control and inter-process communications.  For the malicious activity, powershell.exe and python.exe are present which are indicative of PowerShell Empire (the C2 capability listed in the ground truth).
Furthermore, the lsass.exe process is present which can be associated with benign activity such as logging into a computer; however, frequently the process is co-opted by malicious actors to harvest credentials. 
During malicious activity there is an increase over the normal activity due to the addition of activity without the removal of normal activity. We do note that the OpTC dataset is a test environment (i.e., the normal activity is simulated) and as such the benign user activity threshold may be lower than standard traffic. However, they were simulated to be representative of typical system usage.}

We show in the results that these dynamics of the topology are captured  by the zigzag barcode vectorizations allowing us to detect a difference between benign and malicious activity patterns.


\section{Results} \label{sec:results}

Here we demonstrate the ability of both the zigzag persistence and summary statistics to detect malicious activity for an example source IP. Namely, we demonstrate these results for source IP 142.20.56.202 for malicious activity and source IP 142.20.56.175 for benign activity on host 201 on September 23, 2019. We chose this malicious source IP and host to demonstrate the effectiveness of this autoencoder due to the variety of attacks during this time as shown in the ground truth data provided in the GitHub repository\footnote{See \text{https://github.com/FiveDirections/OpTC-data} for red team ground truth data}. While there are limited malicious source IPs during this time window there are a very large number of benign source IPs. We chose source IP 142.20.56.175 as an exemplary benign source IP, but we found very similar dynamics and reconstruction loss values for other benign source IPs.

\begin{figure*}[p]
    \centering
    \begin{subfigure}[t]{.78\textwidth}
        \centering
        \includegraphics[width=.88\textwidth]{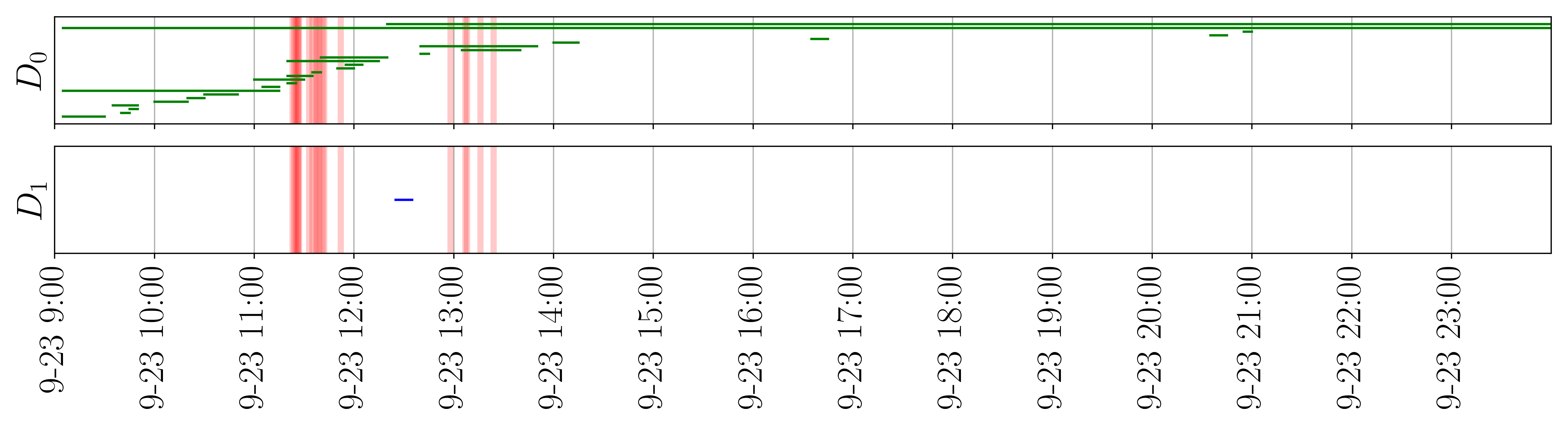}
        \caption{Zigzag persistence barcodes.}
        \label{fig:ZZ_optc_malicious}
    \end{subfigure}
    \vspace{25pt}
    \begin{subfigure}[t]{.78\textwidth}
        \centering
        \includegraphics[width=.99\textwidth]{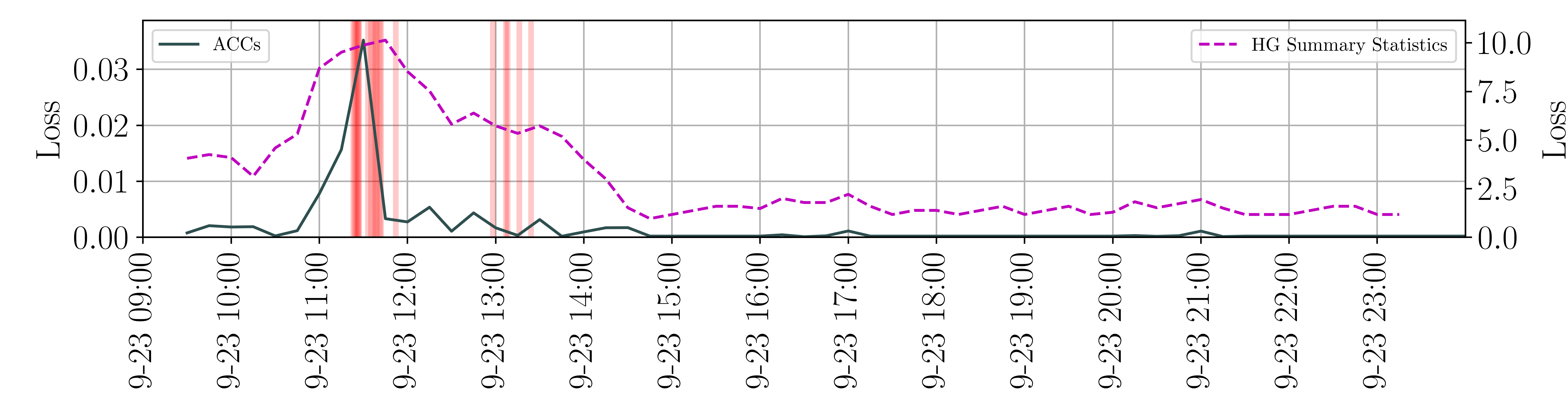}%
        \caption{Autoencoder reconstruction loss (mean squared error).}
        \label{fig:AE_results_malicious}
    \end{subfigure}
    \caption{\chg{Autoencoder results for malicious source IP 142.20.56.202 on host 201 using the ACCs of the windowed zigzag persistence barcode compared to summary statistics with highlight red vertical lines for each malicious activity instance recorded in OpTC ground truth.}}
    \label{fig:malicious_AE_results}
\end{figure*}

Figure~\ref{fig:malicious_AE_results} shows the zigzag persistence barcode (\ref{fig:ZZ_optc_malicious}) and the reconstruction losses over time for the ACC vectors and the hypergraph summary statistics (\ref{fig:AE_results_malicious}). In both plots we have highlighted each of the malicious events from the OpTC ground truth diary as red vertical bars. \chg{From these events it is clear that there are two main sequences of attacks: the first from approximately 11:30 to 12:00 and the second from 13:00 to 13:30. The first group of attacks consist of a password collection attempt through {Mimikatz} to elevate the agent and then attempts at injecting into the LSASS process using {psinject}. The second group of attacks is based on scanning procedures including a {ping sweep} and {ARP scan}.}

The main takeaway from Figure~\ref{fig:malicious_AE_results} is that while both ACCs and hypergraph summary statistics seem to show an anomaly during the malicious activity \chg{through a peak in the reconstruction loss}, the autoencoder trained on the ACCs more precisely detects the \chg{first group of} malicious activity. 
The summary statistics show a broad range in time when the reconstruction loss is high (approximately 9:30 to 14:30) which is larger than the range occupied by the malicious activity. 
On the other hand, the autoencoder trained on the ACCs is able to accurately detect the first sequence of attacks with a \chg{sharp} spike in reconstruction loss from approximately 11:00 to 11:40, which closely correlates to when the first attack sequence occurred.
However, \chg{there is no peak during the second sequence of attacks which was dominated by ARP scans and ping sweeps. 
We believe these were not clearly detected due to the specific hypergraph construction we chose: hyperedges as destination ports and image paths as vertices. 
Our hypergraph construction is not sensitive to this attack as many of the lines in the log data corresponding to ARP scans and ping sweeps are not labeled with a source IP. And when the log data is associated with a source IP they are repetitive (e.g., the ping responses repeatedly have image path System and destination port 0) and do not show up as significant changes in the hypergraph's topology. In future work we plan to use our same pipeline with different hypergraph constructions to better identify different attack types.}

\begin{figure*}[p]
    \centering
    \begin{subfigure}[t]{.78\textwidth}
        \centering
        \includegraphics[width=.86\textwidth]{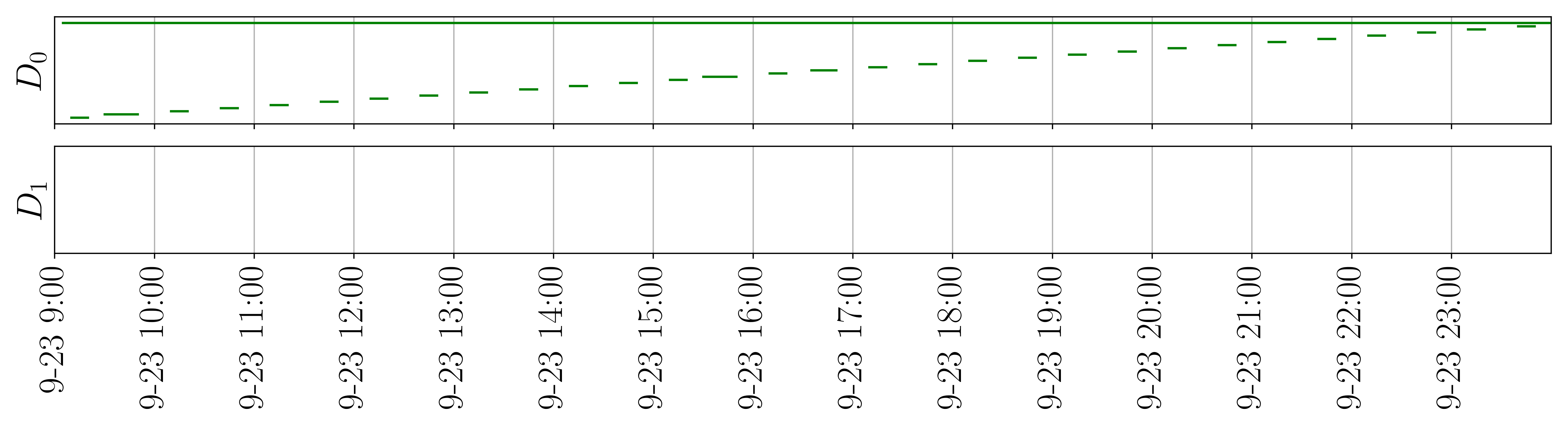}
        \caption{Zigzag persistence barcodes.}
        \label{fig:ZZ_optc_benign}
    \end{subfigure}
    \vspace{25pt}
    \begin{subfigure}[t]{.78\textwidth}
        \centering
        \includegraphics[width=.99\textwidth]{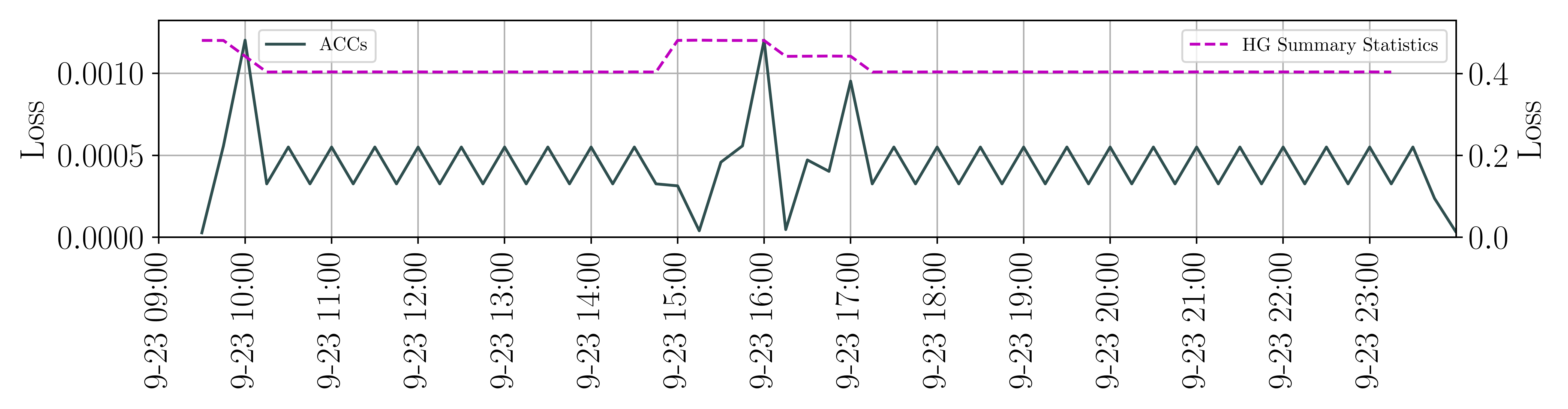}%
        \caption{Autoencoder reconstruction loss (mean squared error).}
        \label{fig:AE_results_benign}
    \end{subfigure}
    \caption{\chg{Autoencoder results for benign source IP 142.20.56.175 on host 201 using the ACCs of the windowed zigzag persistence barcode compared to summary statistics with highlight red vertical lines for each malicious activity instance recorded in OpTC ground truth.}}
    \label{fig:benign_optc_results}
\end{figure*}

As a point of comparison we show the same zigzag and reconstruction loss plots for an exemplary benign source IP in  Fig.~\ref{fig:benign_optc_results}. From the zigzag barcode (\ref{fig:ZZ_optc_benign}) we see that there are typically no 1-dimensional features, as evidenced by the empty $D_1$ barcode, for benign activity. Moreover, the 0-dimensional features have a predictable, periodic behavior. This is further substantiated by the reconstruction loss for both the ACCs and summary statistics being very low (compare the $y$-axis scales in Fig.~\ref{fig:AE_results_benign} to those in Fig.~\ref{fig:AE_results_malicious}).

\chg{To quantify these results across more benign data and demonstrate the consistency we lastly compare the 25th, 50th (median) and 75th percentiles of the distributions of reconstruction losses for both the ACC and summary statistic trained autoencoders during benign and then malicious activity on host 201 on the 23rd as shown in Table~\ref{tab:average_losses}. Additionally, we calculated these same statistics for these autoencoders tested on the training hosts (benign) on the 24th as a point of comparison to the benign activity on the 23rd. By comparing these percentiles we are able to quantitatively confirm the performance of the autoencoders.}
\begin{table}[h]
\chg{\begin{tabular}{lllllll}
\multirow{2}{*}{Host(s)}                   & \multicolumn{3}{c}{ACC ($\times 10^{-3}$)}                                                                & \multicolumn{3}{c}{Summary Statistics}                                                                    \\
                                           & \multicolumn{1}{c}{\textbf{25\%}} & \multicolumn{1}{c}{\textbf{50\%}} & \multicolumn{1}{c}{\textbf{75\%}} & \multicolumn{1}{c}{\textbf{25\%}} & \multicolumn{1}{c}{\textbf{50\%}} & \multicolumn{1}{c}{\textbf{75\%}} \\ \hline
\multicolumn{1}{l|}{201 (Benign IPs)}          & 0.04                              & 0.11                              & \multicolumn{1}{l|}{0.19}         & 0.68                              & 0.92                              & 1.19                              \\
\multicolumn{1}{l|}{201 (Malicious IPs)}       & 1.21                              & 3.93                              & \multicolumn{1}{l|}{7.96}         & 5.31                              & 6.96                              & 8.81                              \\
\multicolumn{1}{l|}{Training Hosts (24th)} & 0.07                              & 0.14                              & \multicolumn{1}{l|}{0.26}         & 0.76                              & 1.03                              & 1.34                              \\ 
\multicolumn{1}{l|}{Training Hosts (23rd)} & 0.06                              & 0.12                              & \multicolumn{1}{l|}{0.26}         & 0.72                              & 1.01                              & 1.39                              \\ \hline
\end{tabular}}
\caption{\chg{The $25^{\rm th}$, $50^{\rm th}$ (median), and $75^{\rm th}$ percentiles of the autoencoder (both ACC and hypergraph summary statistics trained autoencoders) reconstruction loss distributions on host 201 on the 23rd for malicious and benign IP addresses and for all the training hosts on both the 23rd and the 24th.} }

\label{tab:average_losses}
\end{table}


\chg{From Table~\ref{tab:average_losses} it is clear that the interquartile interval from the 25th to the 75th percentiles for host 201 for benign and malicious activity do not overlap for both the ACCs and summary statistics resulting in both autoencoders being able to accurately distinguish between the two states. This is shown with the $75^{\rm th}$ percentile of the benign source IP reconstruction loss being less than (6 times less than) the $25^{\rm th}$ percentile of the malicious source IPs.

We also compare the benign activity on host 201 to the training hosts on both the 23rd and 24th to demonstrate the benign reconstruction loss is similar across hosts and that the autoencoder was not over trained. This  is shown with the autoencoder loss distributions trained on ACCs and summary statistics being similar with all having their inter-quartile intervals significantly overlapping.

Lastly, based on the medians for both the summary statistics and the ACCs it seems the ACCs more clearly detect malicious activity the the median loss being approximately 35.6 times greater during malicious compared to benign activity on host 201 and only 7.5 times greater for the summary statistics.}

\section{Conclusion} \label{sec:conclusion}

The work we present in this paper shows that the dynamics of topology of hypergraphs representing cyber log data can be effective for distinguishing malicious activity from benign.
However, we have noted some limitations that we plan to explore in future work.
In particular, the ACC vectorization strategy for persistence barcodes is rather coarse. We plan to evaluate more complex representations like persistence images and landscapes for this vectorization step. \chg{We additionally plan to study the sensitivity of the autoencoder to both the initialized random weights and the training data (e.g., mixing in some malicious data into training data).}
We are also aware that our hypergraph construction linking executables to destination ports does not capture all types of malicious behavior. We will experiment with additional hypergraph constructions to understand how other malicious behavior can be encoded.
\chg{Along those lines we additionally plan to test our methods on data sets beyond OpTC to ensure generalizability of the approach, and compare to other approaches of studying hypergraph data including hypergraph neural networks.}
Finally, in order for cyber analysts to trust the results of our pipeline we must be able to provide some \chg{interpretation} of the \chg{specific} topological \chg{features in $H_0$ and $H_1$} in the context of the log data and ground truth malicious activity. This is ongoing work and provides an exciting opportunity for collaboration between cybersecurity researchers and mathematicians.




\bibliographystyle{splncs04}
\bibliography{references}


\end{document}